\documentclass{llncs}

\usepackage[utf8]{inputenc} 
\usepackage{graphicx}

\usepackage{dcolumn}
\usepackage[detect-family]{siunitx}
\usepackage{rotating}
\usepackage{booktabs}
\usepackage{tabularx}
\usepackage{newfloat}
\DeclareFloatingEnvironment[
    fileext=loa,
    listname=List of Algorithms,
    name=Algorithm,
    placement=tbhp,
]{algorithm}
\usepackage{algpseudocode}

\usepackage[caption=false]{subfig}

\graphicspath{ {./figures/} }

\title{Structure of a media co-occurrence network}

\author{V.A. Traag\inst{1} \and R. Reinanda\inst{2} \and G. van Klinken\inst{1}}
\institute{%
       KITLV, Leiden, the Netherlands%
  \and Faculty of Science, University of Amsterdam, Amsterdam, the Netherlands%
  }
\begin{document}

\maketitle

\begin{abstract} 

Social networks have been of much interest in recent years. We here focus on a
network structure derived from co-occurrences of people in traditional
newspaper media. We find three clear deviations from what can be expected in a
random graph. First, the average degree in the empirical network is much lower
than expected, and the average weight of a link much higher than expected.
Secondly, high degree nodes attract disproportionately much weight. Thirdly,
relatively much of the weight seems to concentrate between high degree nodes.
We believe this can be explained by the fact that most people tend to co-occur
repeatedly with the same people.  We create a model that replicates these
observations qualitatively  based on two self-reinforcing processes: (1) more
frequently occurring persons are more likely to occur again; and (2) if two
people co-occur frequently, they are more likely to co-occur again. This
suggest that the media tends to focus on people that are already in the news,
and that they reinforce existing co-occurrences.

\end{abstract}

\section{Introduction}

Complex networks have been a prominent research topic for the past decade. One
of the reasons is that complex networks appear in a multitude of scientific
disciplines, varying from
neurology~\cite{bullmore_complex_2009,hagmann_mapping_2008}, ecology
\cite{garlaschelli_universal_2003,guimera_origin_2010} to international
relations~\cite{cranmer_kantian_2014,maoz_what_2008,garlaschelli_structure_2005}
and human mobility~\cite{gonzalez_understanding_2008,simini_universal_2012}
providing a unified theoretical framework for analysis. Although many properties
seem to be (nearly) universal (e.g. degree distribution,
clustering)~\cite{barabasi_scale-free_2009}, there are also some noteworthy
differences between different types of networks (e.g.  assortativity, weak
links)~\cite{petri_topological_2013,guimera_classes_2007,amaral_classes_2000}. 

Social networks can nowadays be relatively easily scraped from online services
such as Facebook or
Twitter~\cite{corten_composition_2012,traud_social_2012,ferrara_large-scale_2012}.
In addition, traditional media (i.e. newspapers) are also increasingly being
digitised. Whereas online social media are open to the general public, and the
large masses use them intensively, traditional media are biased towards the
more influential members of society. Therefore, we might learn something about
the elite of a society by studying how they appear in the newspapers. In this
study, we focus on who co-occurs with whom. We aim to understand the structure
of the co-occurrences. Do frequently occurring people co-occur mostly with each
other, or not? How strongly do frequently occurring people connect to each
other?

Given the possibilities of using media co-occurrence reports for constructing
social networks, there have been surprisingly few earlier studies of media
co-occurrence
networks~\cite{steinberger_cross-lingual_2007,pouliquen_extracting_2008,ozgur_social_2004,joshi_discovering_2006}.
The analyses were relatively succinct, and showed that such networks were both
scale-free and a small world, properties we also find here. However, the
scale-free and small world nature of these co-occurrence networks can be
expected, and also show up in a randomised graph. In fact, various common
properties are quite close to what can be expected in a randomised graph.
Nonetheless, three deviations from the random graph stand out. First of all, the
average degree is much higher than expected, while the average weight is much
lower than expected. Secondly, high degree nodes attract disproportionately much
weight.  Thirdly, much of the weight is concentrated in between high degree
nodes.

These observations suggest that people repeatedly occur with the same people, or
at least more so than expected at random. To explain this, we create a model
that concentrates more of the co-occurrences in fewer people, thus explaining
this deviation. The model consists of only two simple ingredients: (1) more
frequently occurring people tend to occur more frequently in the future; and (2)
people that co-occur more frequently tend to co-occur more frequently in the
future. In addition, high degree nodes are more likely to occur with already
existing neighbours.

\section{Data \& Network}

\begin{sidewaystable}[ph!]
  \caption{Summary overview. We report various properties for both
datasets and the randomisations and model. Refer to the main text for details
about the randomisation procedures and the model.}
  \label{tab:summary}
  \centering
  \sisetup{detect-family, mode=text}
  \begin{tabular}{%
      ll
      D{|}{\,\pm\,}{-1}
      D{|}{\,\pm\,}{-1}
      D{|}{\,\pm\,}{-1}
    }
    \toprule
        \rule{1ex}{0pt} & 
      & \multicolumn{1}{c}{Empirical} 
      & \multicolumn{1}{c}{Random} 
      & \multicolumn{1}{c}{Model} \\
    \midrule
    \multicolumn{5}{l}{Joyo} \\
    & Nodes 
        & \num{9567}  
        & \num{9114} | \num{418.0} 
        & \num{9481} | \num{93.5} \\
    & Avg. Degree
        & \num{12.4}  
        & \num{22.1} | \num{8.0}
        & \num{12.3} | \num{8.8e-2} \\
    & Avg. Weight
        & \num{2.9}
        & \num{1.2} | \num{0.050}
        & \num{3.0} | \num{8.9e-3} \\
    & Avg. Strength
        & \num{36.5}  	
        & \num{27.7} | \num{11.0}
        & \num{36.4} | \num{0.36} \\[2ex]

    & Assortativity 
        & \num{-0.067}
        & \num{-0.13}  | \num{8.3e-3}
        & \num{-0.094} | \num{1.3e-3} \\
    & Clustering		
        & \num{0.29} 
        & \num{0.32}  | \num{2.2e-4}
        & \num{0.25}  | \num{8.2e-4} \\
    & W. Clustering  
        & \num{0.33}
        & \num{0.34}  | \num{2.3e-4}
        & \num{0.26}  | \num{6.9e-4} \\[2ex]

    & Path Length  
        & \num{3.45}
        & \num{2.51} | \num{4.3e-4}
        & \num{3.34} | \num{1.5e-3} \\
    & Diameter  
        & \num{10} 
        & \num{5.7} | \num{4.9e-2}
        & \num{8} | \num{5.8e-2} \\
    & Radius  
        & \num{6} 
        & \num{3} 
        & \num{4.5} \\[2ex]

    & Exponents \\
    & Degree
        & \num{2.46} | \num{1.5e-2} 
        & \num{2.29} | \num{1.4e-2}
        & \num{2.03} | \num{1.1e-2} \\
    & Weight
        & \num{2.21} | \num{1.2e-2}
        & \num{2.56} | \num{1.6e-2}
        & \num{2.45} | \num{6.0e-3} \\
    & Strength
        & \num{2.06} | \num{1.1e-2}
        & \num{2.17} | \num{1.2e-2}
        & \num{1.89} | \num{9.1e-3} \\
    & Degree-Strength
        & \num{1.30} | \num{2.6e-3}
        & \num{1.42} | \num{1.4e-3}
        & \num{1.39} | \num{1.8e-3} \\[4ex]

    \multicolumn{5}{l}{NYT} \\
    & Nodes
        & \num{31093}
        & \num{30860} | \num{69}
        & \num{31015} | \num{11} \\
    & Avg. Degree
        & \num{22.3}
        & \num{45.2} | \num{0.19}
        & \num{22.1} | \num{0.14} \\
    & Avg. Weight
        & \num{2.01}
        & \num{1.11} | \num{6.6e-4}
        & \num{1.94} | \num{0.025} \\
    & Avg. Strength
        & \num{44.9}
        & \num{50.1} | \num{0.24}
        & \num{43.1} | \num{0.27} \\[2ex]

    & Assortativity
        & \num{0.062}
        & \num{-0.090} | \num{6.4e-4}
        & \num{-0.17} | \num{2.4e-3} \\
    & Clustering
        & \num{0.32}
        & \num{0.26} | \num{4.2e-4}
        & \num{0.36} | \num{7.7e-3} \\
    & W. Clustering
        & \num{0.35}
        & \num{0.26} | \num{4.6e-4}
        & \num{0.36} | \num{7.7e-3} \\[2ex]

    & Path Length
        & \num{3.7}
        & \num{2.7} | \num{5.9e-3}
        & \num{3.2} | \num{4.6e-3} \\
    & Diameter
        & \num{10}
        & \num{6}
        & \num{9} \\
    & Radius
        & \num{5}
        & \num{3}
        & \num{4.5} \\[2ex]

    & Exponents \\
    & Degree
        & \num{3.99} | \num{1.7e-2}
        & \num{2.63} | \num{9.3e-3}
        & \num{1.77} | \num{4.4e-3} \\
    & Weight
        & \num{2.42} | \num{8.1e-3}
        & \num{2.90} | \num{1.1e-2}
        & \num{2.71} | \num{2.9e-3} \\
    & Strength
        & \num{2.95} | \num{1.1e-2}
        & \num{2.51} | \num{8.6e-3}
        & \num{1.72} | \num{4.1e-3} \\
    & Degree-Strength
        & \num{1.48} | \num{3.5e-3}
        & \num{1.22} | \num{4.8e-4}
        & \num{1.48} | \num{1.1e-3} \\

    \bottomrule
  \end{tabular}
\end{sidewaystable}

We use newspaper articles to construct a social network. The idea is that people
are linked if they co-occur in the same sentence. We use two corpora in our
current study: (1) a corpus from an Indonesian news service called
\emph{Joyo\footnote{\url{http://www.joyonews.org}}}; and (2) a corpus from the
New York Times\footnote{See \url{https://catalog.ldc.upenn.edu/LDC2008T19} for
the corpus. We only used the first two years of the dataset.} (NYT). The Joyo
dataset covers roughly 2004--2012 and contains $140\,263$ articles, while the
NYT dataset covers 1987--1988 and contains $210\,645$ articles. The Joyo dataset
is a selection of political news (in English) from both domestic and foreign
sources that is relevant to the politics of Indonesia, while the NYT is a
complete corpus of all the articles of that newspaper.  We scan the whole text
and automatically identify entities by using a technique known as named entity
recognition (NER)~\cite{finkel_incorporating_2005}. The technique automatically
identifies different entities, and classifies them into three distinct
categories: persons, organisations and locations. Although it is not perfect,
the error rate tends to be relatively low~\cite{finkel_incorporating_2005}. 

We have only included persons in our media co-occurrence network, and only
people that occurred in more articles than on average. We thereby exclude
people that only appear quite infrequently, which are presumably less
influential, and thus less of interest. Although this skews the results more
towards people that appear more prominently in the media, it still includes
many less prominent people. Once all persons have been identified, we have to
disambiguate them. There are generally two types of errors that can be made
with names~\cite{milne_learning_2008}: (1) a single name corresponds to two
different persons (e.g. ``Bush'' can refer to the $43^\text{rd}$ or
$41^\text{st}$ US president); and (2) two different names refer to the same
person (``President Clinton'' or ``Bill Clinton'' both refer to the
$42^\text{nd}$ US president). The second problem appears much more prominent
than the first problem in our corpus, as people are generally referred to in
many different ways in journalistic prose (including or not positions, titles,
initials, maiden names, etc\ldots). 

We disambiguated these names by using a combination of similarity measures based
on Wikipedia matching, string similarity and network similarity (using Jaccard
similarity). The more prominent people often have a Wikipedia page, and various
spelling variants are redirected to the same entity (e.g. ``President Clinton''
and ``Bill Clinton'' both redirect to the same Wikipedia page). Each similarity
is normalised to fall between $0$ and $1$ (with $1$ being identical), which we
threshold at $0.75$, such that we only take it into account if the similarity is
at least $0.75$. We then take the average of the similarities that are higher
than $0.75$ (which is then also at least $0.75$). We then find clusters of names
such that each cluster has an average internal similarity of $0.85$ (all
similarity measures are between $0$ and $1$), using a technique called the
Constant Potts Model~\cite{traag_narrow_2011}.

Once the names have been disambiguated, we create a link for all the unique
names in a sentence (i.e. repeated names have no effect). We do this for every
sentence, and simply count in how many sentences such a co-occurrence was
observed. We take only the largest connected component of the network, and this
constitutes the media co-occurrence network we analyse in this paper.

Of course, what co-occurrence exactly implies is not always clear: two people
might be mentioned together for example because they collaborate, or because
they are contestants in an election. A co-occurrence might not coincide with any one single
definition of a ``relationship'' in the sociological
sense~\cite{knoke_social_2007}. Hence, we cannot say if two people that
co-occur have any more significant relationship: do they know each other? Have
they ever communicated?  Have they met face to face? Are they close friends?
Sworn enemies? We simply cannot tell. This is essential to bear in mind when
drawing any conclusions: the network is based on co-occurrence, not on ``actual
relationships''.

\begin{figure}[tbh!]
  \begin{center}
    \includegraphics[width=\textwidth]{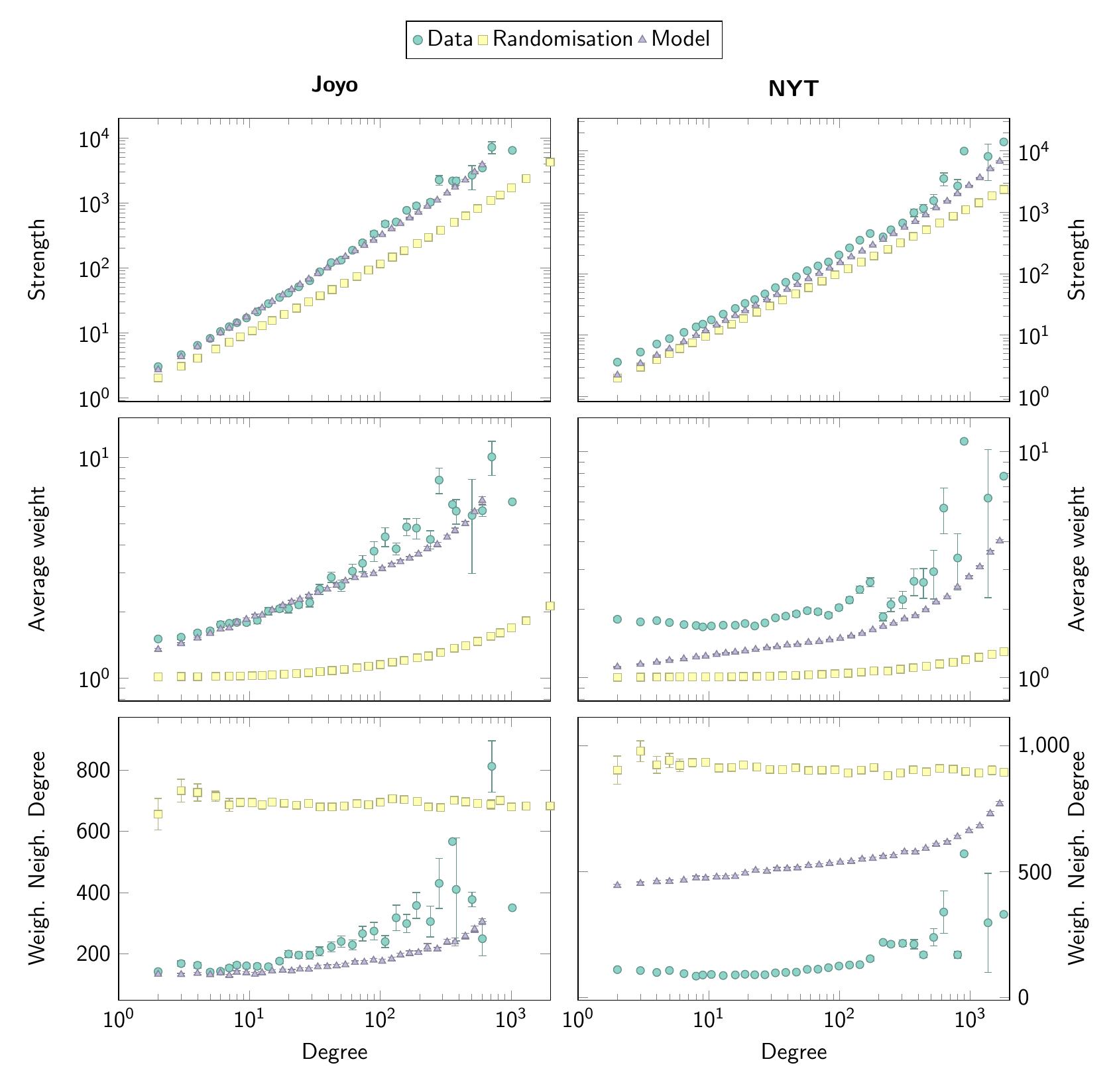}
  \end{center}
  \caption{Properties. The first column shows the properties for Joyo,
  the second for NYT. This clearly shows that the strength increases faster than
  expected (first row), which is even more clear when looking at the average
  strength (second row). The model however, captures quite well this increase,
  especially for Joyo. The weighted neighbour degree increases, whereas this
  remains nearly constant for the randomisation.  The model again, shows a
  relatively similar increase, especially for Joyo.
}
  \label{fig:props}
\end{figure}

\section{Results}

We denote the undirected network by $G=(V,E)$ where $V=\{1,\ldots,n\}$ is
the node set, representing the people, and $E \subseteq V \times V$ constitutes
the edge set with $m = |E|$ edges, representing the co-occurrences between
people. Hence, if node $i \in V$ and node $j \in V$ co-occurred in some
sentence, then there is an edge $(ij) \in E$. Each edge has a weight associated
to it, which represents the number of times the two nodes $i$ and $j$ have
co-occurred, which we denote by $w_{ij}$. Finally, the adjacency matrix is
denoted by $A$, such that $A_{ij} = 1$ if there is an edge between node $i$ and
$j$ and zero otherwise. Since the network is undirected, we have that $A_{ij} =
A_{ji}$ and $w_{ij} = w_{ji}$. Notice that this network is a projection of a
bipartite network of people and sentences. Let us denote by $B$ the bipartite
adjacency matrix, so that $B_{is} = 1$ if person $i$ occurs in sentence $s$.
Then $w_{ij} = \sum_s B_{is}B_{js}$ and $A_{ij} = 1$ if $w_{ij} > 0$ and $A_{ij}
= 0$ otherwise. In other words, $w = BB^T$. We denote the degree of node $i$ by
$k_i = \sum_j A_{ij}$ and the strength by $s_i = \sum_j A_{ij}w_{ij}$.

We compare our results to a bipartite randomisation of the network. Formally,
let $P$ be a list of persons and $S$ a list of sentences, such that $(P_e S_e)$
is a bipartite edge, so that the length of the list equals the number of
bipartite edges. If $\sigma$ is a random permutation, then $(P_{\sigma_e} S_e)$
for all $e$ are the randomized edges. Simply put, this randomisation takes the
list of occurrences of people in sentences and shuffles the complete edge list,
so that people occur in a random sentence. This preserves the number of times
somebody occurs in a sentence, and preserves the number of people that occur in
a sentence. We then take a projection of this network as we did for the
empirical observations. In other words, $\hat{w} = \hat{B}\hat{B}^T$ and
$\hat{A}_{ij} = 1$ if $\hat{w}_{ij} > 0$, where $\hat{B}_{P_{\sigma_e} S_e} = 1$
for all $e$ and zero otherwise. We create a $100$ different randomisations and
use it to compare to our empirical observations.

The Joyo network has $n=9\,467$ nodes, an average degree of $\langle k_i \rangle
\approx 12.22$ and with an average weight per edge of $\langle w_{ij} \rangle
\approx 2.95$ the average strength is $\langle s_i \rangle = \langle k_i \rangle
\langle w_{ij} \rangle \approx 36.07$. So, in Joyo, a person co-occurs on
average about $3$ times with $12$ other people, so in total about $36$ times.
Based on the randomisation, we would expect roughly $22$ people to occur about
$1.2$ times, giving a total strength of around $28$. Hence, people co-occur
roughly $2.5$ times more often with the same people as expected, and co-occur
with almost $2$ times less people.  The NYT corpus contains $n=31\,093$ nodes
and has an average degree of about $\langle k_i \rangle \approx 22.35$. The
average weight $\langle w_{ij} \rangle \approx 2.01$, which gives an average
strength of about $\langle s_i \rangle \approx 44.91$.  In summary, a person in
the NYT co-occurs about $2$ times with about $22$ different people. Similar as
for Joyo, based on the randomisation, we would expect around $45$ people to
occur about $1.1$ times, giving a total strength of about $50$. So, in NYT
people co-occur almost $2$ times more frequently with the same people, with
roughly $2$ times fewer people. We provide an overview of some of the key
statistics in Table~\ref{tab:summary}.

The degree, strength and weight are all heterogeneously distributed, as
frequently observed in complex networks. They follow approximately
powerlaws~\cite{clauset_power-law_2009} in both networks, where we use MLE
techniques for estimating the powerlaws. The degree and strength in the Joyo
corpus is more broadly distributed compared to the NYT corpus, but the weight is
distributed similarly. However, these distributions are also expected to be
heterogeneous based on the randomisation. Although there are some deviations,
they seem mainly due to the lower average empirical degree and the higher
average empirical weight. 

The network shows signs of a small world network. The average path
length is relatively low, while the clustering is relatively high. Again, this
does not deviate much from what is expected at random. The (weighted) clustering
is almost the same, and the path length is even slightly longer than expected at
random. The longer path length is probably due to the lower average degree. With
fewer neighbours on average, there are fewer possibilities for paths, thus
leading to somewhat longer paths.

We find that the strength scales superlinearly with the degree as $s_i \sim
k_i^{\beta}$, with an exponent of $\beta \approx 1.30$ for Joyo and $\beta
\approx 1.48$ for NYT. This means that high degree nodes attract more weight and
low degree nodes less weight. In this context, people that co-occur with many
other people, also tend to co-occur more often. A similar superlinear scaling
was found in transportation and technological
networks~\cite{wang_general_2005,ou_power-law_2007,barthelemy_characterization_2005}.
This contrasts with for example mobile phones where there is a sublinear
growth~\cite{onnela_analysis_2007}, suggesting that people who call many people,
do so less frequently than people who call few people. This is quite different
from what is expected at random, especially when looking at this from the
perspective of the average weight $s_i/k_i$, which increases for large degree
$k_i$ empirically, but which increases only very slightly in the randomisation.
We have plotted the behaviour of the average weight and the strength in
Fig.~\ref{fig:props}.

The average weighted neighbour degree~\cite{barrat_architecture_2004} increases
with larger degree. This implies that high degree nodes connect relatively
stronger to other high degree nodes. For the NYT the weighted neighbour degree
starts to increases more clearly for a degree larger than about $200$. This
suggest that relatively much weight is in between high degree nodes. See
Fig.~\ref{fig:props} for plots of the weighted neighbour degree. The ordinary
neighbour degree decreases for Joyo, as evidenced by the negative
assortativity~\cite{Newman2002}, whereas this increases for NYT
(Table~\ref{tab:summary}), pointing out a difference between the two datasets.

\subsection{Model}

Three phenomena deviate from what can be expected from a random graph. First,
the degree is higher than expected and the weight is lower than expected.
Secondly, high degree nodes attract disproportionately much weight.  Thirdly,
much of the weight is between the hubs. These observations suggest that people
tend to co-occur repeatedly with the same people. We therefore introduce a very
simple stylistic model that is able to reproduce most of the observations in the
empirical network qualitatively. The model consists of two key ingredients: (1)
more frequently occurring people have a higher probability of occurring; and (2)
two more frequently co-occurring people have a higher probability of
co-occurring. 

More specifically, we employ the following procedure. We start out with an empty
graph. Each time step, we draw a random sentence $s$, with degree $k_s$ (i.e.
the number of people occurring in a sentence) drawn from the empirical sentence
degree distribution. We then choose $k_s$ nodes in the following way. With
probability $q$ we introduce a new person into the graph, which is chosen so
that the expected number of nodes equals the number of nodes $n$ in the
empirical graph. That is, if $p_k$ is the probability a sentence has degree $k$,
then the total expected number of nodes occurring in sentences will be $n_s
\sum_k k p_k$, with $n_s$ the number of sentences. So, if $q = \frac{n}{n_s
\sum_k k p_k}$ we generate on average about $n$ nodes.

If we don't introduce a new node, we pick a random node $i$ in the sentence.
Then with probability $(k_i + 1)^{-\beta}$ we choose an already existing node,
where $k_i$ is the degree of node $i$ and $\beta$ a tunable parameter. The
probability a node is selected is proportional to its degree, so that
$\Pr(\text{choose~} j) = k_j / \sum_l k_l$.  If we don't pick an existing node,
we choose a random neighbour of $i$ with probability proportional to the weight,
so that $\Pr(\text{choose~} j | i) = w_{ij}/\sum_k w_{ik}$. After we picked all
$k_s$ nodes, we create an edge for all combinations of persons in the sentence.
If an edge already exists, we increase its weight.

The node sampling is very similar to the preferential attachment model from
Barabási and Albert \cite{barabasi_emergence_1999} and similar
models~\cite{Dorogovtsev2000a}, as nodes that have more links are more likely to
receive additional links. However, our model differs in several important ways
from the model by Barabási and Albert \cite{barabasi_emergence_1999}.  First of
all, it tends to generate a superlinear scaling of the strength with the degree.
Secondly, it generates a much higher clustering coefficient.  This latter effect
is mainly a result of sampling neighbours, which relates to triadic closure,
which has also been used in other models~\cite{Kumpula2007}.

Besides preferential attachment, our model also has a counter tendency. Higher
degree nodes are increasingly more likely to co-occur with already existing
neighbours. The idea behind the scaling $(k_i + 1)^{-\beta}$ is based on the
idea that higher degree nodes have a higher than linear strength.  In other
words, hubs are more likely to repeatedly co-occur with their neighbours, more
so than on average.

A crude argument shows that indeed this model should result in superlinear
scaling of the strength. Consider $k_i(s_i)$, the degree of node $i$ as a
function of the strength $s_i$, and suppose that all sentences have only degree
$2$. Every time we add a co-occurrence for $i$, $s_i$ increases, although
$k_i$ does not necessarily increase. Now if $i$ was the first node to be chosen,
it will get a new neighbour with probability $(k_i + 1)^{-\beta}$. If $i$ was the second node
to be chosen, it implies it is chosen by another node. The probability that this
node selected a new neighbour (which by definition then is node $i$, since we
already know it was chosen) is then $(k_j + 1)^{-\beta}$ given that node $j$ was
chosen first. But the probability that node $j$ was chosen first is
proportional to $k_j$. Then the probability $i$ gets a new neighbour if $s_i$
increases is 
\begin{equation}
  (k_i + 1)^{-\beta} + \sum_j \frac{k_j}{\sum_l k_l} (k_j + 1)^{-\beta}.
\end{equation}
Taking a mean-field approach, we approximate $k_i \approx \langle
k \rangle$, and simply write $k = \langle k \rangle$ for ease of writing, we
obtain that
\begin{equation}
  \Delta k = (k + 1)^{-\beta} + \sum_j \frac{k}{\sum_l k} (k + 1)^{-\beta} =
  2(k+1)^{-\beta}.
\end{equation}
Taking then the approximation that $\partial k / \partial s \approx \Delta k$,
we obtain the solution that $s \sim k^{1 + \beta}$ so that the strength
increases superlinearly with $k$. This is of course a rather crude argument, but
it nonetheless shows that using this approach we should indeed expect a
superlinear scaling of the strength with the degree. 

This contrasts with using a constant probability for choosing a new neighbour.
In that case, essentially each time that the strength increases, the probability
that the degree increases is a fixed probability $\rho$. This then results in $k
\sim \rho s$, showing only linear scaling.

Additionally, this model has the tendency to create a relatively high clustering
coefficient. Every time that a new sentence is introduced with $k_s$ persons,
all these $k_s$ persons will be connected amongst each other, creating small
cliques. In addition, these cliques are also reinforced by the mechanism of
adding neighbours to sentences. Interestingly, despite such reinforcement, the
model still generates a dissortative structure~\cite{Newman2002}. Although this
is congruent with Joyo, it contrasts with the assortative structure for NYT.

\begin{figure}[tbh!]
  \begin{center}
    \includegraphics{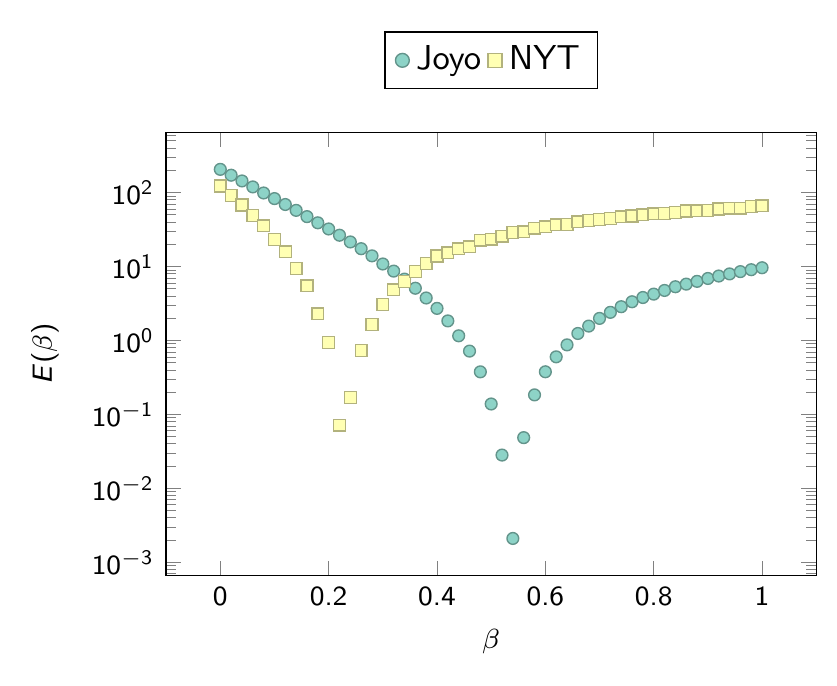} \\
  \end{center}
  \caption{Model fit. The fit of the parameter $\beta$ in the model.
The error $E(\beta)$ it taken with respect to the average degree and the average
weight.}
  \label{fig:parameter_fit}
\end{figure}

In order to estimate the parameter $\beta$, we compare the average degree and
average weight to the empirically observed values. That is, for each parameter
value $\beta$, we compare the average degree $\langle \hat{k}_i \rangle$ and
average weight $\langle \hat{w}_{ij} \rangle$ of the model, and compare that to
the average degree $\langle k_i \rangle$  and average weight $\langle w_{ij}
\rangle$ of the empirical network. We use a simple squared error for the
fitting, 
\begin{equation}
  E(\beta) = (\langle \hat{k}_i \rangle - \langle k_i \rangle)^2 + 
          (\langle \hat{w}_{ij} \rangle - \langle w_{ij} \rangle)^2.
\end{equation}
We used $100$ replications for each parameter value. The best parameter fit
differs quite a bit between Joyo and NYT. For Joyo we find an optimal parameter
of $\beta \approx 0.46$, while for NYT we find $\beta \approx 0.22$. See
Fig.~\ref{fig:parameter_fit} for the fitting of the parameter.

As expected, we can fit the empirically observed average degree and weight very
well. Additionally, we indeed observe a very similar increase in the average
weight for high degree nodes, as was already argued above. Finally, the model
also shows an increase in the average weighted neighbour degree, similar as
empirically. Nonetheless, although the average degree is quite well fit for
NYT, the distribution of the degree is more broader, leading to a higher average
weighted neighour degree. Nonetheless, both the model and the empirical network
show an increase. The fit of the model for Joyo is especially striking in
Fig.~\ref{fig:props}. Perhaps this is because Joyo has a more particular focus
on politics, while the NYT includes also other subjects such as culture, arts
and sports.

\section{Conclusion}

In this paper we analysed two networks based on the co-occurrence of people in
newspapers. We have analysed various properties of this network, and whereas
many properties are in line with what could be expected from such a
co-occurrence network, a few deviations stand out. First, people occur with
fewer people than expected and more often with those people than expected.
Secondly, high degree nodes attract disproportionately much weight, so that the
hubs co-occur much more often than their degree justifies. Third, much of the
weight concentrates between these hubs. 

This suggests that people repeatedly co-occur with the same people. We
constructed a model that tries to reproduce these observations. It is based on
two simple processes: (1) people that occur in the media are more likely to
occur again; (2) two people that co-occur are more likely to co-occur again.
Moreover, people with a higher degree co-occur more often with people with whom
they already co-occur. This seems to explain the observations quite well,
although some deviations remain.

There are some clear differences between the Joyo and the NYT corpus.  Whether
this is reflective of differences between Indonesia and the US, or the more
politically oriented corpus of Joyo, or a difference in time periods, is
difficult to ascertain. Further analysis and comparison of these networks should
provide more insight.

\section*{Acknowledgements}
  VT would like to thank Fabien Tarissan for interesting comments and remarks on
  an earlier version of this manuscript. This research is funded by the Royal
  Netherlands Academy of Arts and Sciences (KNAW) through its eHumanities
  project~\footnote{\url{http://www.ehumanities.nl/computational-humanities/elite-network-shifts/}}.

\bibliographystyle{splncs03}
\bibliography{bibliography_formatted}      

\end{document}